\documentclass[conference, 9pt]{IEEEtran}
\IEEEoverridecommandlockouts
\usepackage{cite}
\usepackage{amsmath,amssymb,amsfonts}
\usepackage{algorithmic}
\usepackage{graphicx}
\usepackage{textcomp}
\usepackage{xcolor}
\def\BibTeX{{\rm B\kern-.05em{\sc i\kern-.025em b}\kern-.08em
		T\kern-.1667em\lower.7ex\hbox{E}\kern-.125emX}}
\usepackage{amsmath}
\usepackage{booktabs} 
\usepackage{multirow}
\usepackage{xcolor}
\usepackage[super]{nth}
\usepackage{tikz}
\newcommand*\circled[1]{\tikz[baseline=(char.base)]{
		\node[shape=circle,draw,inner sep=0.2pt] (char) {#1};}}
\usepackage{soul}
\usepackage{enumitem}
\usepackage{algorithm}

\usepackage{fancyhdr}
\fancypagestyle{firstpage}{
  \fancyhf{}
  \chead{To appear at the 57th Design Automation Conference (DAC), July 2020, San Francisco, CA, USA.}
  \cfoot{\thepage}
}

\begin{document}

\title{\huge DRMap: A Generic DRAM Data Mapping Policy for\\ Energy-Efficient Processing of Convolutional Neural Networks
}
	
\author{\IEEEauthorblockN{Rachmad Vidya Wicaksana Putra, Muhammad Abdullah Hanif, Muhammad Shafique}
\IEEEauthorblockA{\textit{Technische Universit\"at Wien}, Vienna, Austria \\
Email: \{rachmad.putra, muhammad.hanif, muhammad.shafique\}@tuwien.ac.at}
}

\maketitle
\thispagestyle{firstpage}

\begin{abstract}

Many convolutional neural network (CNN) accelerators face performance- and energy-efficiency challenges which are crucial for embedded implementations, due to high DRAM access latency and energy. Recently, some DRAM architectures have been proposed to exploit subarray-level parallelism for decreasing the access latency. Towards this, we present a design space exploration methodology to study the latency and energy of different mapping policies on different DRAM architectures, and identify the pareto-optimal design choices. The results show that the energy-efficient DRAM accesses can be achieved by a mapping policy that orderly prioritizes to maximize the row buffer hits, bank- and subarray-level parallelism. 
\end{abstract}

\begin{IEEEkeywords}
DRAM mapping, DRAM architectures, subarray-level parallelism, convolutional neural networks, CNNs, CNN accelerators. 
\end{IEEEkeywords}

\section{Introduction}

The widespread use of machine learning (ML) algorithms for organizing, analyzing, and inferring information from digital data is growing fast. 
Among many ML algorithms, convolutional neural network (CNN) algorithms have demonstrated state-of-the-art performance in data analytic tasks, such as image classification, object recognition, smart environment, health care, and automotive \cite{Ref:LeCun_DeepLearning_Nature15}. 
Since the CNN algorithms require data-intensive processing, CNN hardware accelerators are typically required to expedite the inference process. 
Over the past few years, several CNN accelerators have been proposed \cite{Ref:Chen_DianNao_ASPLOS14,Ref:Zhang_CNNfpga_FPGA15, Ref:Chen_Eyeriss_JSSC16, Ref:Albericio_Cnvlutin_ISCA16, Ref:Luo_DaDianNao_TC17, Ref:Jouppi_TPU_ISCA17, Ref:Parashar_SCNN_ISCA17, Ref:Lu_FlexFlow_HPCA17, Ref:Kwon_MAERI_ASPLOS18, Ref:Hanif_MPNA_arXiv18, Ref:Sharma_BitFusion_ISCA18}. 
These accelerators offer higher performance- and energy-efficiency as compared to general-purpose CPUs. 
However, many CNN accelerators still face performance- and energy-efficiency challenges due to the high off-chip memory (i.e., DRAM) access latency and energy, 
which are higher than the latency and energy for the other compute operations \cite{Ref:Sze_DNNsurvey_IEEE17}.
\textit{Therefore, reducing the DRAM access latency and energy are required for improving the performance- and energy-efficiency of CNN accelerators.}

\subsection{The State-of-the-Art and Limitations}
\label{Sec:SoA}

Previous works have proposed different techniques to reduce the DRAM access energy, by minimizing the number of DRAM accesses \cite{Ref:Zhang_CNNfpga_FPGA15}\cite{Ref:Li_SmartShuttle_DATE18}\cite{Ref:Stoutchinin_Scheduling_arXiv19}.
Their main ideas are similar, i.e., (i) defining layer partitioning\footnote{\scriptsize Layer partitioning determines portions of data in the form of block/tile to be accessed from DRAM to on-chip memory at one time. 
A detailed explanation is provided in Section \ref{Sec:PartitionScheduleCNNs}.} and then transferring each partition from DRAM to on-chip memory/buffer in a defined schedule, and (ii) maximally reusing the data that are already in the on-chip buffer. 
The state-of-the-art \cite{Ref:Li_SmartShuttle_DATE18} considers adaptive layer partitioning and scheduling, to minimize the number of DRAM accesses, by adaptively switching the reuse priority between different data types: input activations/feature maps (\textit{ifms}), output activations/feature maps (\textit{ofms}), and weights (\textit{wghs}), across the layers of a network. 
\textit{Although all these works result in a reduced number of DRAM accesses (which also means reduced DRAM access energy), they do not consider improving (i) the DRAM latency-per-access, and (ii) DRAM energy-per-access. 
Therefore, performance- and energy-efficiency improvements achieved by the state-of-the-art are sub-optimal, thereby limiting the CNN accelerators to achieve further performance- and energy-efficiency improvements.}
We will illustrate this with the help of the following motivational case study.

\subsection{Motivational Case Study and Associated Research Challenges}

\begin{figure}[hbtp]
  \centering
  \vspace{-0.2cm}
  \includegraphics[width=\linewidth]{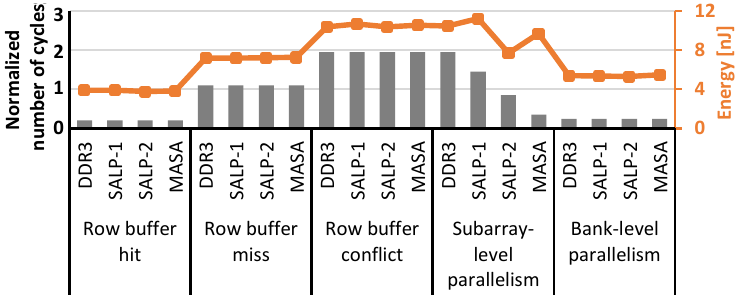}
  \vspace{-0.6cm}
  \caption{DRAM latency-per-access and DRAM energy-per-access for different conditions (a row buffer hit, a row buffer miss, a row buffer conflict, subarray- and bank-level paralellism) in different DRAM architectures (DDR3, SALP-1, SALP-2, and SALP-MASA). Data are obtained from our experiments using a state-of-the-art cycle-accurate DRAM simulators \cite{Ref:Kim_Ramulator_LCA15}\cite{Ref:Ghose_VAMPIRE_POMACS18} for DDR3-1600 2Gb x8 and SALP 2Gb x8 with 8 subarrays-per-bank.}
  \label{Fig:DRAM_HitMissConflict}
\end{figure}

\textbf{Motivational Case Study:} Although there are various types of commodity DRAM (e.g., DDR3, DDR4, etc.), they have similar internal organization 
and operations 
\cite{Ref:Ghose_DRAMworkload_Sigmetrics19} (detailed DRAM organization and operations are provided in Section \ref{Sec:DRAMintro}). 
Therefore, different types of commodity DRAM have similar behavior regarding latency-per-access and energy-per-access.
%
The DRAM latency-per-access and energy-per-access vary depending upon whether a single DRAM access faces \textit{a row buffer hit}, \textit{a row buffer miss}, or \textit{a row buffer conflict}.
A row buffer hit means that the requested row is already available in the row buffer, hence the data access can be performed directly without additional operations.
In case of a row buffer miss or conflict, the requested row has to be opened first before a data access can be performed.
In this manner, a row buffer miss and conflict require higher latency-per-access and energy-per-access than a row buffer hit.
To illustrate this, we performed an experimental analysis to observe the DRAM latency-per-access and energy-per-access for different conditions (i.e., a row buffer hit, miss, and conflict), and the experimental results are presented in Fig. \ref{Fig:DRAM_HitMissConflict}. 
%
Furthermore, in commodity DRAM, each request that goes to a DRAM bank can only access a single DRAM subarray at a time, although each bank is composed of multiple subarrays.
This limits the DRAM capability to offer lower DRAM access latency and energy. 
Recently, several DRAM architectures that offer subarray-level parallelism (SALP) in a DRAM bank, have been proposed in the literature. In \cite{Ref:Kim_SALP_ISCA12}, three variants of SALP architectures are presented, i.e., SALP-1, SALP-2, and SALP-MASA (detailed SALP architectures are provided in Section \ref{Sec:DRAM_SALP}).
Our observation results in Fig. \ref{Fig:DRAM_HitMissConflict} show that SALP architectures have the potential to further reduce the DRAM latency-per-access and energy-per-access as compared to commodity DRAM. 

\textbf{Associated Research Challenges:} From above observations, the energy efficiency of DRAM accesses for CNN accelerators can be improved by minimizing the DRAM latency-per-access and energy-per-access. 
\textit{Therefore, there is a need of a generic DRAM mapping policy that can achieve maximum row buffer hits while exploiting subarray- and bank-level parallelism.}
Furthermore, to justify that the proposed DRAM mapping policy is applicable to different design choices,
\textit{a design space exploration (DSE) is required}. 
This DSE explores different DRAM mapping policies in different DRAM architectures with different layer partitioning and scheduling schemes, to find the minimum energy-delay-product (EDP) of DRAM accesses. 
This EDP is used as a measure of the energy-efficiency of a CNN accelerator.
Therefore, \textit{an analytical model for estimating the EDP of different DRAM mapping policies in the DSE, is also required}.

\subsection{Our Novel Contributions}

\begin{figure}[hbtp]
  \vspace{-0.3cm}
  \centering
  \includegraphics[width=\linewidth]{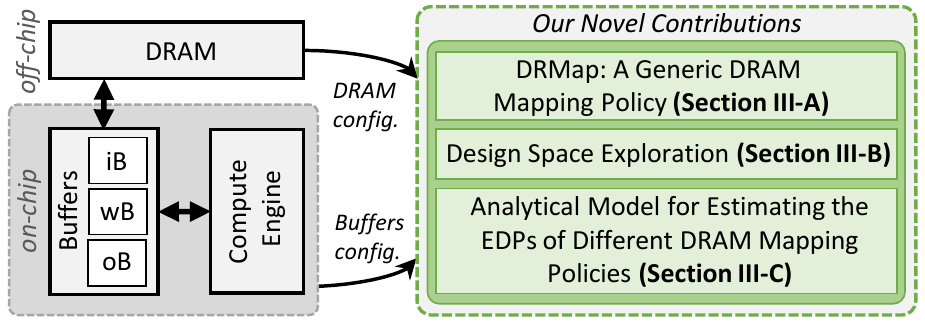}
  \vspace{-0.6cm}
  \caption{The overview of our novel contributions, highlighted in the green boxes. We use separate on-chip buffers for different data types: input buffer (\textit{iB}) for \textit{ifms}, weight buffer (\textit{wB}) for \textit{wghs}, and output buffer (\textit{oB}) for \textit{ofms}.}
  \label{Fig:NovelContributions}
  \vspace{-0.2cm}
\end{figure}

In this paper, we make the following novel contributions (the overview is illustrated in Fig. \ref{Fig:NovelContributions}) to overcome the associated challenges. 

\begin{enumerate}[leftmargin=*]
  \item We propose \textbf{DRMap: a generic \underline{DR}AM data \underline{Map}ping policy} that offers minimum energy-delay-product (EDP) of DRAM accesses, for a given DRAM architecture, layer partitioning and scheduling scheme.
  DRMap orderly prioritizes to maximize row buffer hits, bank- and subarray-level parallelism.
  \item We propose \textbf{a design space exploration (DSE) algorithm} to find a DRAM mapping that offers minimum EDP, while considering different DRAM architectures, different layer partitioning and scheduling schemes.
  \item We propose \textbf{an analytical model for estimating EDPs of different DRAM mapping policies}, which will be used in the DSE. 
  The EDP for each DRAM mapping is estimated by multiplying the number of DRAM accesses with the respective number of cycles and energy values. 
\end{enumerate}

\textbf{Key results:} 
DRMap orderly prioritizes to maximize the row buffer hits, bank- and subarray-level parallelism.
It improves the EDP compared to the other mapping policies, up to 96\% for DDR3, 94\% for SALP-1, 91\% for SALP-2, and 80\% for SALP-MASA on AlexNet \cite{Ref:Alex_AlexNet_NIPS12}.


\renewcommand{\headrulewidth}{0pt}
\section{Preliminaries}
\label{Sec:Prelim}

\subsection{Layer Partitioning and Scheduling in CNNs}
\label{Sec:PartitionScheduleCNNs}

\begin{figure}[hbtp]
  \centering
  \vspace{-0.3cm}
  \includegraphics[width=\linewidth]{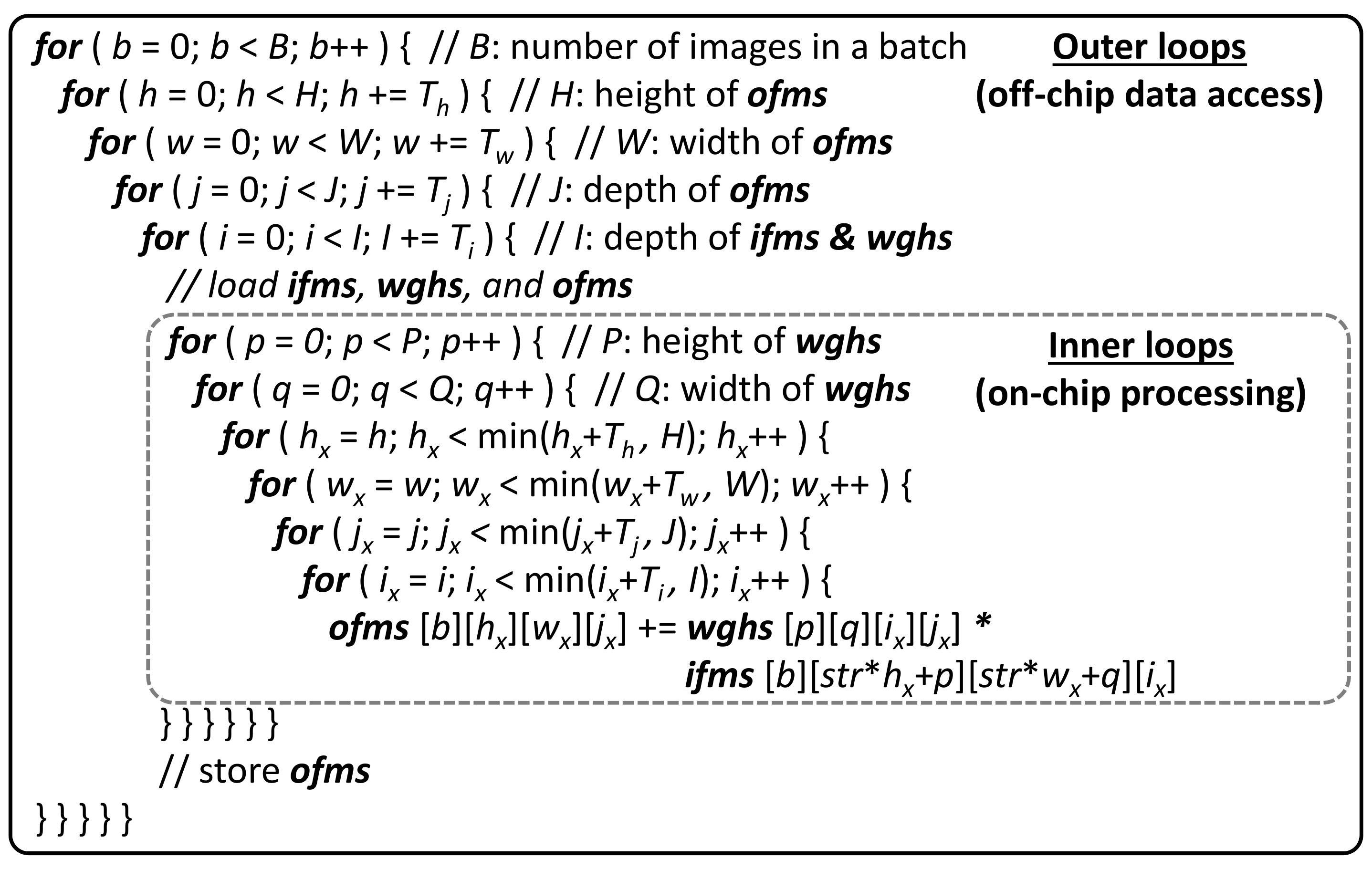}
  \vspace{-0.6cm}
  \caption{Pseudo-code of the tiled convolutional neural network processing.}
  \label{Fig:PseudoCode_CNN}
\end{figure}

The full CNN processing usually cannot be mapped at once on the accelerator fabric due to the limited on-chip buffer capacity (i.e., $100$KB-$500$KB \cite{Ref:Sze_DNNsurvey_IEEE17}), hence layer partitioning and scheduling are required.
To illustrate this, a pseudo-code of a convolutional layer processing in a CNN accelerator is shown in Fig. \ref{Fig:PseudoCode_CNN}.
It has two parts, i.e., inner loops and outer loops.
The inner loops represent the on-chip processing.
The outer loops represent the scheduling of processing different portions of data (from all data types: \textit{ifms}, \textit{wghs}, and \textit{ofms}), whose sizes have to be less than or equal to the sizes of respective buffers (\textit{iB}, \textit{wB}, and \textit{oB}).
These data are partitioned in the form of blocks/tiles which are represented with the step sizes.
Furthermore, the sequence of the outer loops represents the order in which the tiles are accessed from DRAM to the on-chip buffer. 
It thereby reflects the number of DRAM accesses required to process a layer of a network.

\vspace{-0.1cm}
\subsection{DRAM Overview}
\label{Sec:DRAMintro}

\begin{figure}[hbtp]
  \centering
  \vspace{-0.4cm}
  \includegraphics[width=\linewidth]{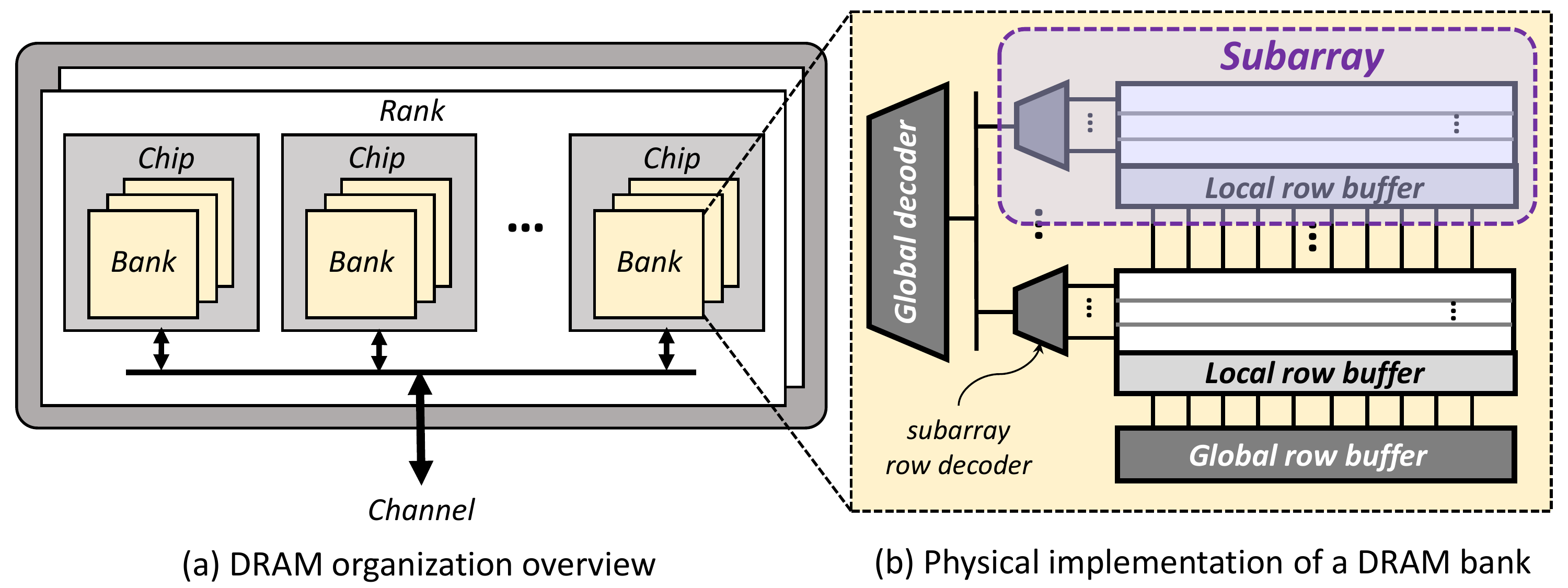}
  \vspace{-0.6cm}
  \caption{(a) DRAM organization overview, and (b) physical implementation of a DRAM bank, showing multiple subarrays in a bank.}
  \label{Fig:DRAMorg}
  \vspace{-0.2cm}
\end{figure}

\textbf{DRAM Organization:} From top to bottom perspective, the organization of a commodity DRAM is composed of \textit{channel}, \textit{rank}, \textit{chip}, \textit{bank}, \textit{row}, and \textit{column} \cite{Ref:Ghose_DRAMworkload_Sigmetrics19}\cite{Ref:Liu_RAIDR_ISCA12}, as shown in Fig. \ref{Fig:DRAMorg}(a). 
In commodity DRAM, banks are the lowest hierarchy in DRAM, which can be accessed in parallel, and referred to as \textit{bank-level parallelism} \cite{Ref:Kim_MemScheduling_MM11}.
Actually, a DRAM bank is \textit{not} implemented in a monolithic design (a large array of cells with a single row buffer). 
Instead, it is implemented in multiple \textit{subarrays}, each of which has its local row buffer, as shown in Fig \ref{Fig:DRAMorg}(b). 
Multiple subarrays in a bank share (i) global bitlines, which connect local row buffers to a global row buffer, and (ii) a global row address decoder \cite{Ref:Kim_SALP_ISCA12}.

\textbf{DRAM Operations:} If there is a single DRAM request, a rank will respond and thereby multiple chips within this rank can be accessed in parallel, contributing to a DRAM word.
For each chip, the request is routed to a specific bank and decoded into addresses of a row and a column.
The \textit{Activation} (ACT) command triggers a row activation, and data from the requested row are copied to the row buffer.
Afterward, a \textit{read} (RD) or \textit{write} (WR) command can be issued to a specific column in the activated row buffer.
If the requested row is already activated, then the data in this row is already in the row buffer (\textit{a row buffer hit}).  
Therefore, it does not need a new row activation.
If the requested row is not activated yet, then it is either \textit{a row buffer miss} or \textit{a row buffer conflict}. 
In \textit{a row buffer miss}, there is no activated row yet in the row buffer and it requires to activate the requested row.
Meanwhile, in \textit{a row buffer conflict}, there is an activated row in the row buffer, but it is not the one that the request is expecting. 
Therefore, this condition requires to close the activated row first using the \textit{precharging} (PRE) command, and then activate the requested row using the \textit{activation} (ACT) command.

\textbf{DRAM Data Mapping:} The default data mapping prioritizes to map the subsequently accessed data in the different columns of the same row of a bank (for increasing row buffer hits) and the different banks of the same rank (for exploiting bank-level parallelism). 
However, it does not exploit subarray-level parallelism and does not consider different possible layer partitioning and scheduling. Therefore, the default data mapping solution is suboptimal. 

\subsection{DRAM Architectures that Exploit Subarray-level Parallelism}
\label{Sec:DRAM_SALP}

In a commodity DRAM, each request that goes to a DRAM bank, can only access a single subarray at a time. 
This limits the potential to reduce the DRAM access latency and energy.
To address this limitation, \cite{Ref:Kim_SALP_ISCA12} has proposed three DRAM architectures and mechanisms that exploit subarray-level paralellism (SALP) in the same bank, called SALP-1, SALP-2, and SALP-MASA. 
Following are the key ideas of these SALP architectures.

\begin{itemize}[leftmargin=*]
    \item \textbf{SALP-1} reduces the DRAM service time by overlapping the \textit{precharging} of one subarray with the \textit{activation} of another subarray, since mostly the \textit{precharging} and \textit{activation} are local to a subarray. To enable this mechanism, re-interpretation of the existing timing constraint for \textit{precharging}, is required.
    \item \textbf{SALP-2} reduces the DRAM service time even more than SALP-1, by overlapping the \textit{write-recovery} latency of an active subarray, with the \textit{activation} of another subarray. To enable this, additional circuitry to activate two subarrays at the same time is required.
    \item \textbf{Multitude of Activated Subarrays (MASA)} reduces the DRAM service time even more than SALP-2, by activating multiple subarrays at the same time (the \textit{activations} of different subarrays are overlapped). To enable this, additional circuitry (more than SALP-2) to activate multiple subarrays at the same time is required.
\end{itemize}

\section{Our Design Methodology for DRAM Mapping in CNN Accelerators}
\label{Sec:DesignMethod}

\subsection{DRMap: A Generic DRAM Data Mapping Policy}
\label{Sec:ProposedDRAMmap}

\begin{figure}[hbtp]
  \vspace{-0.3cm}
  \centering
  \includegraphics[width=\linewidth]{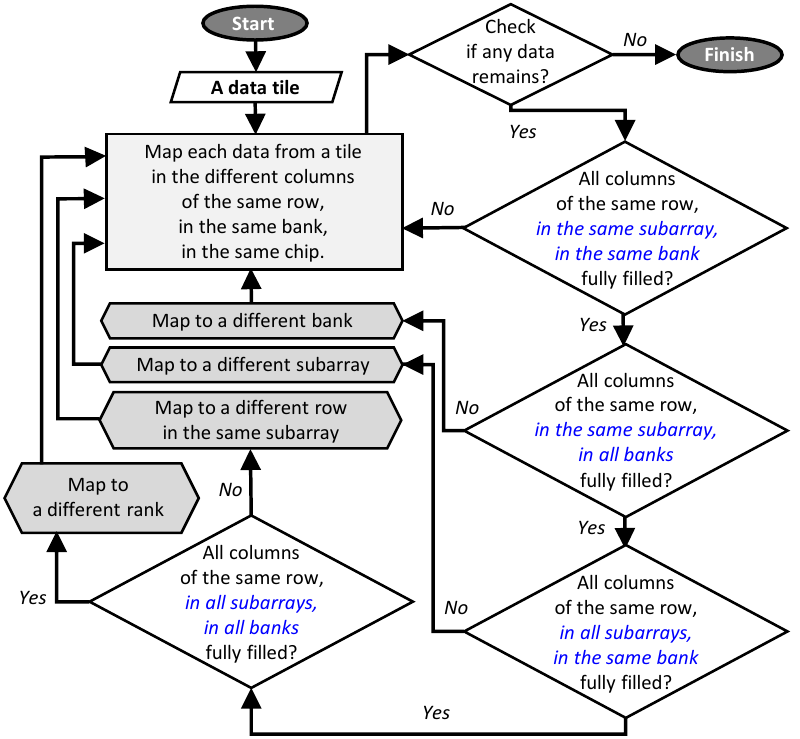}
  \vspace{-0.6cm}
  \caption{Flowchart of the DRMap that illustrates how the mapping policy is performed in a DRAM chip.}
  \label{Fig:DRAMmapFlowchart}
  \vspace{-0.2cm}
\end{figure}

\begin{figure}[hbtp]
  \vspace{-0.3cm}
  \centering
  \includegraphics[width=\linewidth]{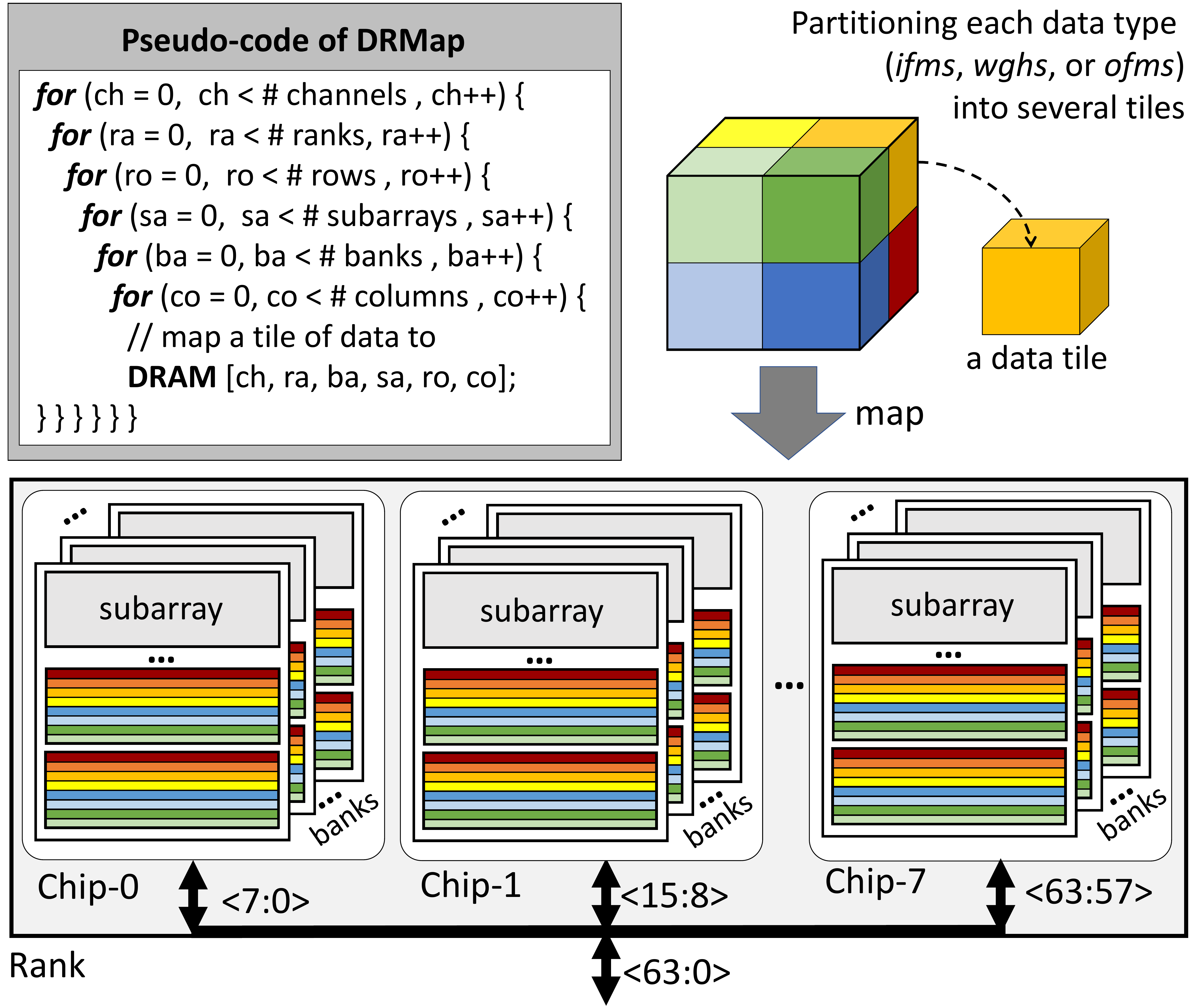}
  \vspace{-0.7cm}
  \caption{Pseudo-code of DRMap and its conceptual implementation in DRAM. 
  }
  \label{Fig:DRMapPhysical}
  \vspace{-0.2cm}
\end{figure}

Our observations from the results in Fig. \ref{Fig:DRAM_HitMissConflict} show that different DRAM architectures have similar behavior in terms of latency-per-access and energy-per-access. 
Therefore, we propose DRMap, a generic DRAM mapping policy for energy-efficient DRAM accesses in CNN accelerators. 
Its main idea is to orderly prioritize the data mapping that maximizes DRAM row buffer hit, bank- and subarray-level parallelism.
The flowchart of DRMap mechanism in a DRAM chip is presented in Fig. \ref{Fig:DRAMmapFlowchart}, while its pseudo-code and physical representation of mapping policy are illustrated in Fig. \ref{Fig:DRMapPhysical}. 
\textit{DRMap considers tile-based partitioning in its mechanism, thereby DRMap can be performed for each data tile using the following steps:}
\begin{enumerate}[leftmargin=*]
    \item If we consider accessing a DRAM bank, then DRMap prioritizes to map a data partition to different columns in the same row to achieve maximum row buffer hits.
    If multiple chips are available within a rank, then this step can be performed in different chips for exploiting the chip-level parallelism.
    \item If all columns in the same row of a bank are fully filled, then the remaining data are mapped to different banks in the same chip, to exploit bank-level parallelism.
    If multiple chips are available, then this step can be performed in different chips. 
    \item If all columns in the same row of all banks are fully filled, then the remaining data are mapped to a different subarray in the same bank, to exploit subarray-level parallelism.
    If multiple chips are available, then this step can be performed in different chips.
    \item If there are remaining data left, then step 1) to 3) can be performed again for different subarray, until all data are mapped within the same rank.
    In this manner, DRMap can achieve maximum row buffer hits, while maximally exploiting bank- and subarray-level parallelism within a DRAM rank.
    \item If there are remaining data left, they can be mapped in different rank (channel) if available, using the same steps as 1) to 4).
    In this manner, our DRMap can achieve maximum row buffer hits, while maximally exploiting bank- and subarray-level parallelism in another DRAM rank (channel) as well.
\end{enumerate}

To illustrate that our DRMap always achieves the minimum energy-delay-product (EDP) of DRAM accesses in different possible conditions, we perform an extensive design space exploration (DSE). The DSE investigates different DRAM mapping policies, different DRAM architectures, as well as different layer partitioning and scheduling schemes on CNN, and estimates EDP for these different combinations. \textit{This DSE is important to corroborate that the best solution that provides the minimum EDP in each given combination is always the same as provided by our DRMap technique.}

\subsection{Design Space Exploration for Evaluating Different DRAM Mapping Policies}
\label{Sec:MethodDSE}

\begin{figure}[hbtp]
  \vspace{-0.4cm}
  \centering
  \includegraphics[width=\linewidth]{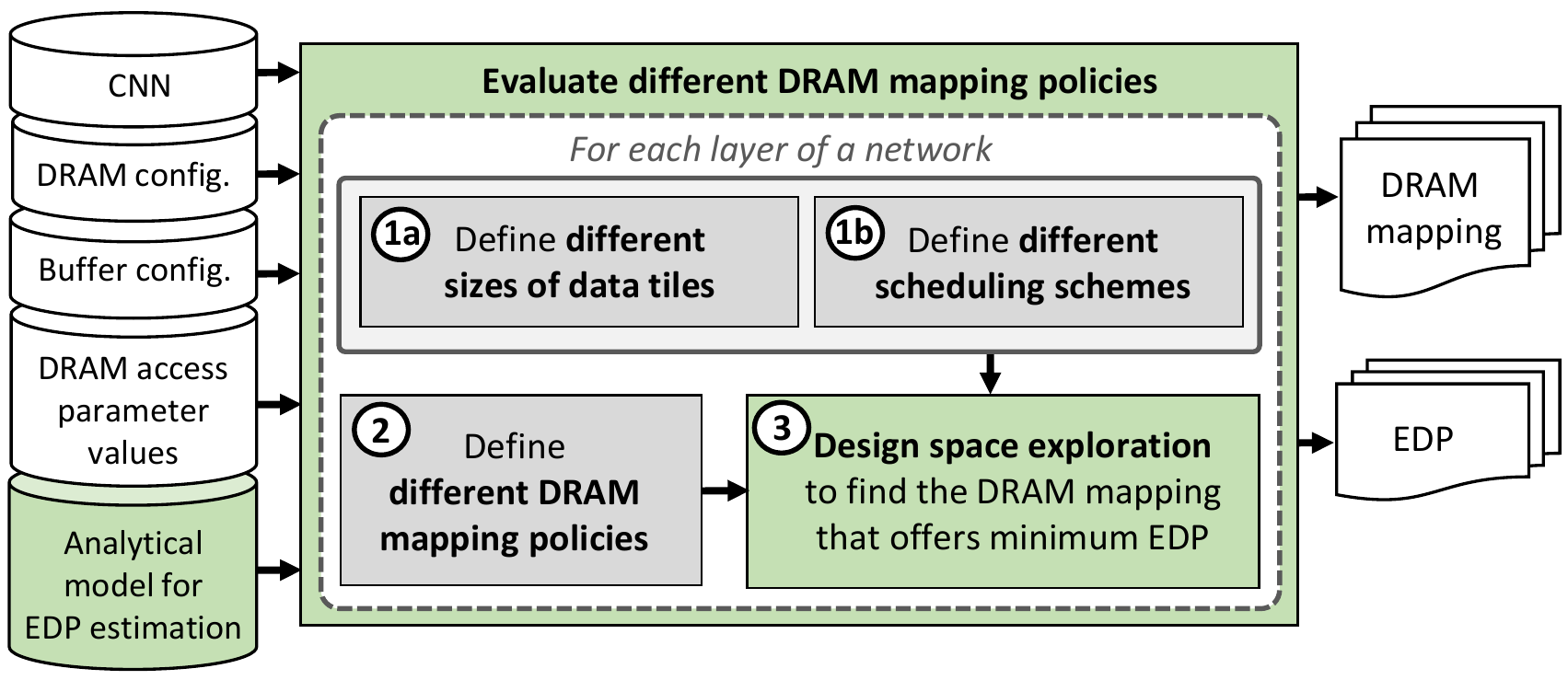}
  \vspace{-0.7cm}
  \caption{The operational flow of the our DSE methodology. Our contributions are highlighted in the green boxes.}
  \label{Fig:MethodOverview}
  \vspace{-0.4cm}
\end{figure}

To evaluate the impact of different DRAM mapping policies and see the performance of DRMap as compared to others, we performed an extensive design space exploration (DSE). 
An overview of the DSE is shown in Fig. \ref{Fig:MethodOverview} and its algorithm is presented in Algorithm \ref{Alg:DSE}. 

\begin{algorithm}
  \caption{Pseudo-code of the proposed DSE algorithm}
  \label{Alg:DSE}
  \begin{algorithmic}[1]
    \renewcommand{\algorithmicrequire}{\textbf{INPUT:}}
	\renewcommand{\algorithmicensure}{\textbf{OUTPUT:}}
	\REQUIRE \textbf{(1)} CNN configuration: number of layers ($L$);\\ 
	\textbf{(2)} Buffer size: \textit{ifms} (\textit{iB}), \textit{wghs} (\textit{wB}), \textit{ofms} (\textit{oB}); \\
	\textbf{(3)} Analytical models of EDP ($EDP$); \\ 
	\textbf{(4)} Layer partitioning for \textit{ifms}, \textit{wghs}, and \textit{ofms} ($Partitioning$); \\ 
	\textbf{(5)} DRAM access scheduling ($Scheduling$); \\ 
	\textbf{(6)} DRAM mapping policies ($DRAMmaps$); \\ 
	\ENSURE \textbf{(1)} Efficient DRAM mapping ($map$); \\
	\textbf{(2)} Minimum EDP ($minEDP$); \\
	\textbf{BEGIN} \\
	  \textbf{Initialization}: \\
	  \STATE $T_p = P$; \\
	  \STATE $T_q = Q$; \\
	  \STATE $EDP[] = 0$; \\
	  \STATE $minEDP[] = 0$; \\
	  \textbf{Process}: \\
		\FOR{($l = 1$ to $L$)}
		  \FOR{(each $Partitioning$)}
		    \FOR{(each $Scheduling$)}
		      \FOR{(each $DRAMmaps$)}
		        \IF{(\textit{ifms} tile size $\leq iB$) \AND (\textit{wghs} tile size $\leq wB$) \AND (\textit{ofms} tile size $\leq oB$)}
		          \STATE Calculate $EDP[l]$;\\
		          \IF{(first loop)}
		            \STATE $minEDP[l] = EDP[l]$;\\
		          \ELSIF{($EDP[l] \leq minEDP[l]$)}
		            \STATE $minEDP[l] = EDP[l]$;\\
		            \STATE Save $map$, $minEDP$;\\
		          \ENDIF\\
		        \ENDIF\\
		      \ENDFOR\\
		    \ENDFOR\\
		  \ENDFOR\\
		\ENDFOR\\
		\RETURN (1) $map$; (2) $minEDP$; \\
	\textbf{END}
	\end{algorithmic} 
\end{algorithm}
\vspace{-0.2cm}
For each layer of a network, the DSE performs three key steps: (1) defining different sizes of data tiles and scheduling schemes, (2) defining different DRAM mapping policies, (3) performing the DSE to find a DRAM mapping policy that offers minimum EDP.
The operational flow of the DSE is explained in the following points:

\textbf{Step-\circled{1}.} Define \circled{1a} different sizes of data tiles for all data types (\textit{ifms}, \textit{wghs}, and \textit{ofms}), and \circled{1b} different scheduling schemes. 
The tile sizes are determined by the step sizes in the outer loops of the Fig. \ref{Fig:PseudoCode_CNN}. 
The tile sizes of \textit{ifms}, \textit{wghs}, and \textit{ofms} have to fit in the corresponding buffers ( \textit{iB}, \textit{wB}, and \textit{oB}). 
\textit{Each combination of the tile sizes for all data types defines one possible partitioning, which will be considered in the DSE.}
The scheduling schemes are determined by the sequence of the outer loops of the Fig. \ref{Fig:PseudoCode_CNN}.
In this work, we consider four scheduling schemes, based on the reuse priority of the data type: \textit{ifms-reuse}, \textit{wghs-reuse}, \textit{ofms-reuse}, and \textit{adaptive-reuse scheduling schemes}. 
The \textit{ifms-reuse} scheduling means that \textit{ifms} data type will be maximally reused when the data are available in the on-chip buffer.
Similar definition is also applied for \textit{wghs-reuse} and \textit{ofms-reuse}.
Meanwhile, the \textit{adaptive-reuse} scheduling means that the reuse priority changes across different layers of a network, according to which one among \textit{ifms}-/\textit{wghs}-/\textit{ofms-reuse} scheduling  that offers minimum number of DRAM accesses.

\textbf{Step-\circled{2}.} Define different DRAM mapping policies, by determining the different orders of mapping loops to different columns, rows, subarrays, and banks in the same DRAM chip. 
For DDR3, orders of mapping loops are permutation of banks, rows, and columns, in the same DRAM chip; meanwhile for SALP, orders of mapping loops are permutation of banks, subarrays, rows, and columns, in the same DRAM chip. 
Here, we narrow down the design space by selecting the DRAM mapping policies that have the least frequent subsequent accesses to different \textit{rows}, since it is the most expensive access in the same DRAM chip, for both latency and energy (as validated by Fig. \ref{Fig:DRAM_HitMissConflict}).
Therefore, there are six mapping policies to be explored in the DSE, as presented in Table \ref{Table:DRAMmaps}.

\textbf{Step-\circled{3}.} Perform the DSE to find a DRAM mapping policy that offers minimum EDP, across different DRAM architectures, different layer partitioning and scheduling schemes.
The minimum EDP and the corresponding DRAM mapping are the outputs of the DSE, for a given DRAM architecture, layer partitioning and scheduling. 

Note that the time and energy are already included in the DSE, for determining the EDP in the final results. 
For each layer of a network, EDP is obtained by multiplying the DRAM access energy and latency consumed by each combination of different DRAM mapping policies, different DRAM architectures, as well as different sizes of layer partitioning and scheduling schemes. 
Therefore, DSE will be able to find the combination that offers minimum EDP for each layer of a network and minimum total EDP for a whole network.

\begin{table}[hbtp]
  \vspace{-0.2cm}
  \caption{Different DRAM mapping policies for the DSE.}
  \vspace{-0.1cm}
  \label{Table:DRAMmaps}
  \centering
  \small
  \begin{tabular}{|c|l|}
  \hline 
  \textbf{Mapping} & \textbf{Inner-most- to outer-most-loops} \\
  \hline
  \hline 
  1 & column, subarray, bank, row \\
  \hline
  2 & subarray, column, bank, row \\
  \hline
  3 & column, bank, subarray, row \\
  \hline
  4 & bank, column, subarray, row \\
  \hline
  5 & subarray, bank, column, row \\
  \hline
  6 & bank, subarray, column, row \\
  \hline
  \end{tabular}
  \vspace{-0.3cm}
\end{table}


\vspace{-0.1cm}
\subsection{Analytical Model of Energy-Delay-Product (EDP) Estimation for Different DRAM Mapping Policies}
\label{Sec:AnalyticalModel}

Based on the proposed DSE, \textit{the optimization problem is formulated to minimize the EDP of DRAM accesses for each layer of a network} and can be stated as
  \begin{equation}
    \small
    \begin{split}
      Objective:& \, minimize \, (EDP_{layer})
    \label{Eq:OptimProblemLayer}
    \end{split}
  \end{equation}
The EDP-per-layer ($EDP_{layer}$) is obtained by multiplying the energy-per-layer and latency-per-layer. The energy-per-layer is obtained by accumulating all access energy values incurred from the DRAM accesses for all data tiles.
The latency-per-layer is obtained by accumulating all access latency values incurred from the DRAM accesses for all data tiles.
The access latency and energy are calculated on the basis of DRAM accesses for each data tile since we consider layer partitioning approach. 
Therefore, for each tile, the number of cycles required for DRAM accesses can be formulated as Eq. \ref{Eq:Ncycle_1tile} and the DRAM access energy can be formulated as Eq. \ref{Eq:Energy_1tile}. 
%
\begin{equation}
 \small
  \begin{split}
    Ncycle_{tile} = & Naccess_{dif\_column} \cdot Ncycle_{dif\_column} + \\ 
	& Naccess_{dif\_rows} \cdot Ncycle_{dif\_rows} + \\
	& Naccess_{dif\_subarrays} \cdot Ncycle_{dif\_subarrays} + \\
	& Naccess_{dif\_banks} \cdot Ncycle_{dif\_banks}
	\label{Eq:Ncycle_1tile}
	\end{split}
	\end{equation}
	\begin{equation}
	\small
	\begin{split}
	E_{tile} = & Naccess_{dif\_column} \cdot E_{dif\_column} + \\ 
	& Naccess_{dif\_rows} \cdot E_{dif\_rows} + \\
	& Naccess_{dif\_subarrays} \cdot E_{dif\_subarrays} + \\
	& Naccess_{dif\_banks} \cdot E_{dif\_banks}
	\label{Eq:Energy_1tile}
	\end{split}
\end{equation}
Term $Naccess_{dif\_x}$ denotes the number of accesses to different DRAM-$x$. 
$Ncycle_{dif\_x}$ denotes the number of cycles incurred when accessing different DRAM-$x$. 
$E_{dif\_x}$ denotes the access energy incurred when accessing different DRAM-$x$. For all terms, $x \in$ \{columns, rows, subarrays, banks\}.

\section{Evaluation Methodology}
\label{Sec:EvalMethod}

\begin{figure}[hbtp]
    \vspace{-0.3cm}
	\centering
	\includegraphics[width=\linewidth]{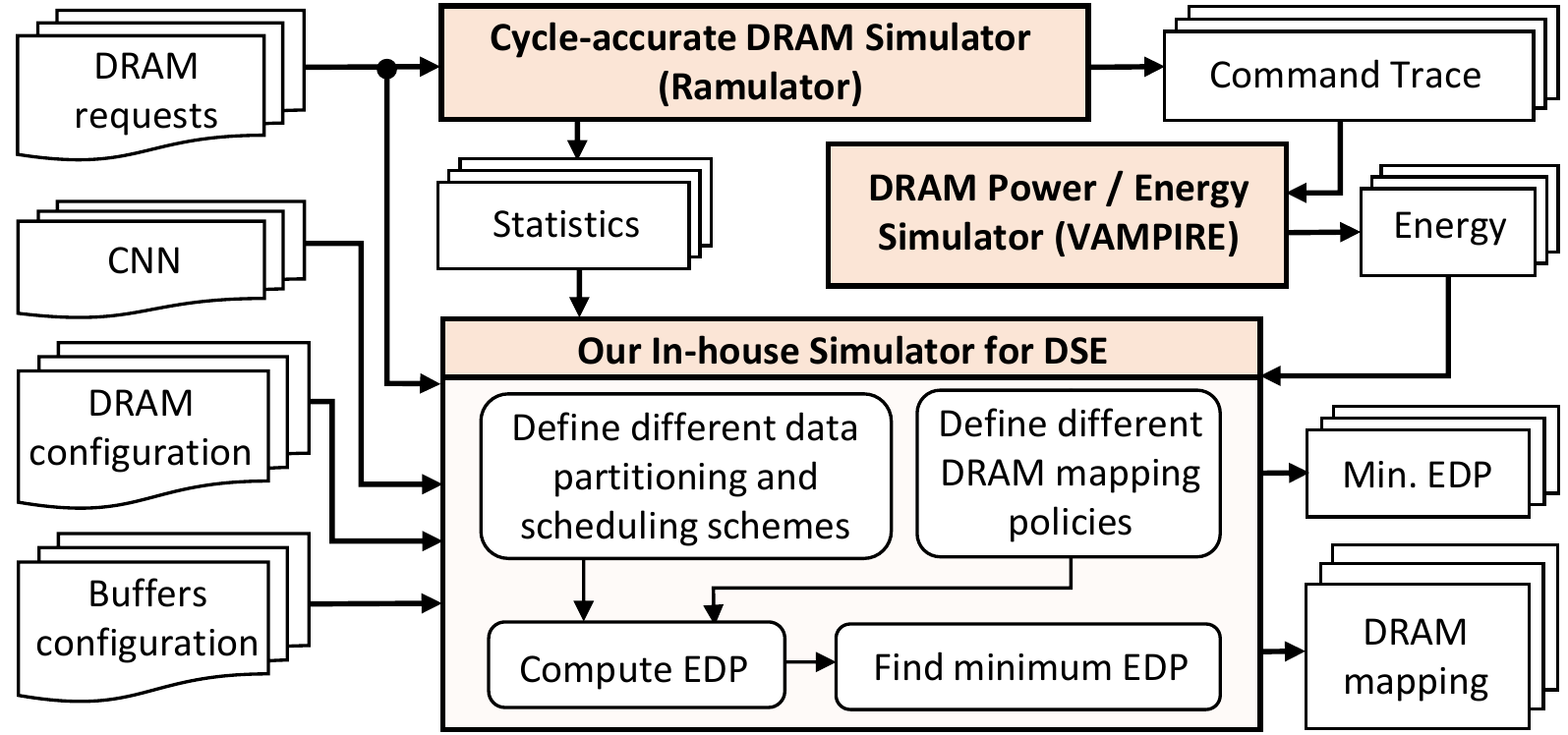}
	\vspace{-0.7cm}
	\caption{Experimental setup and tool flow.}
	\label{Fig:ExpSetup_ToolFlow}
	\vspace{-0.2cm}
\end{figure}

To evaluate our proposed methodology, we built the experimental setup, as presented in Fig. \ref{Fig:ExpSetup_ToolFlow}.

\begin{table}[hbtp]
\vspace{-0.3cm}
\caption{Configuration of the CNN accelerator.}
\vspace{-0.2cm}
\label{Table:Accelerator}
\centering
\small
\begin{tabular}{|l|l|}
\hline 
\multicolumn{1}{|c|}{\textbf{Module}} & \multicolumn{1}{c|}{\textbf{Description}} \\
\hline
\hline
CNN Processing Array & Size = 8 $\times$ 8 MACs \\
\hline
On-chip Buffers & \textit{iB}: 64KB, \textit{wB}: 64KB, \textit{oB}: 64KB \\
\hline
\multirow{1}{*}{Memory Controller} &  Policy = open row, scheduler = FCFS\\
\hline
\multirow{3}{*}{DDR3-1600} & Configuration: 2Gb x8\\ 
     & 1 channel, 1 rank-per-channel, \\
     & 1 chip-per-rank, 8 banks-per-chip \\
\hline
\multirow{4}{*}{SALP} & Configuration: 2Gb x8\\ 
     & 1 channel, 1 rank-per-channel, \\
     & 1 chip-per-rank, 8 banks-per-chip, \\
     & 8 subarrays-per-bank \\
\hline
\end{tabular}
\vspace{-0.2cm}
\end{table}

\textbf{Tool flow:} We used a cycle-accurate DRAM simulator, \textit{Ramulator} \cite{Ref:Kim_Ramulator_LCA15}, to obtain the statistics (i.e, number of cycles) of different DRAM access conditions: a row buffer hit, row buffer miss, row buffer conflict, subarray- and bank-level paralellism.
To profile the energy, we used a real experiments-based DRAM energy simulator, \textit{VAMPIRE} \cite{Ref:Ghose_VAMPIRE_POMACS18}.
Information of energy and number of cycles are used for the DSE, which considers different DRAM mapping policies, different DRAM architectures, different layer partitioning and scheduling schemes to find the DRAM mapping policy that offers minimum EDP.
For DSE, we considered a state-of-the-art Tensor Processing Unit (TPU)\cite{Ref:Jouppi_TPU_ISCA17}-like CNN accelerator with a reduced size of on-chip buffers and MAC array engine, as specified in Table \ref{Table:Accelerator}.
To represent different DRAM architectures, we used DDR3 and SALP architectures (SALP-1, SALP-2, and SALP-MASA). 
For scheduling, we considered \textit{ifms-reuse}, \textit{wghs-reuse}, \textit{ofms-reuse}, and \textit{adaptive-reuse} scheduling schemes.
For mapping, we considered the six mapping policies presented in Table \ref{Table:DRAMmaps}.
For the input, we used AlexNet \cite{Ref:Alex_AlexNet_NIPS12} with ImageNet dataset.

\section{Results and Discussions}
\label{Sec:Results}

\begin{figure*}[t]
    \vspace{-0.3cm}
	\centering
	\includegraphics[width=\linewidth]{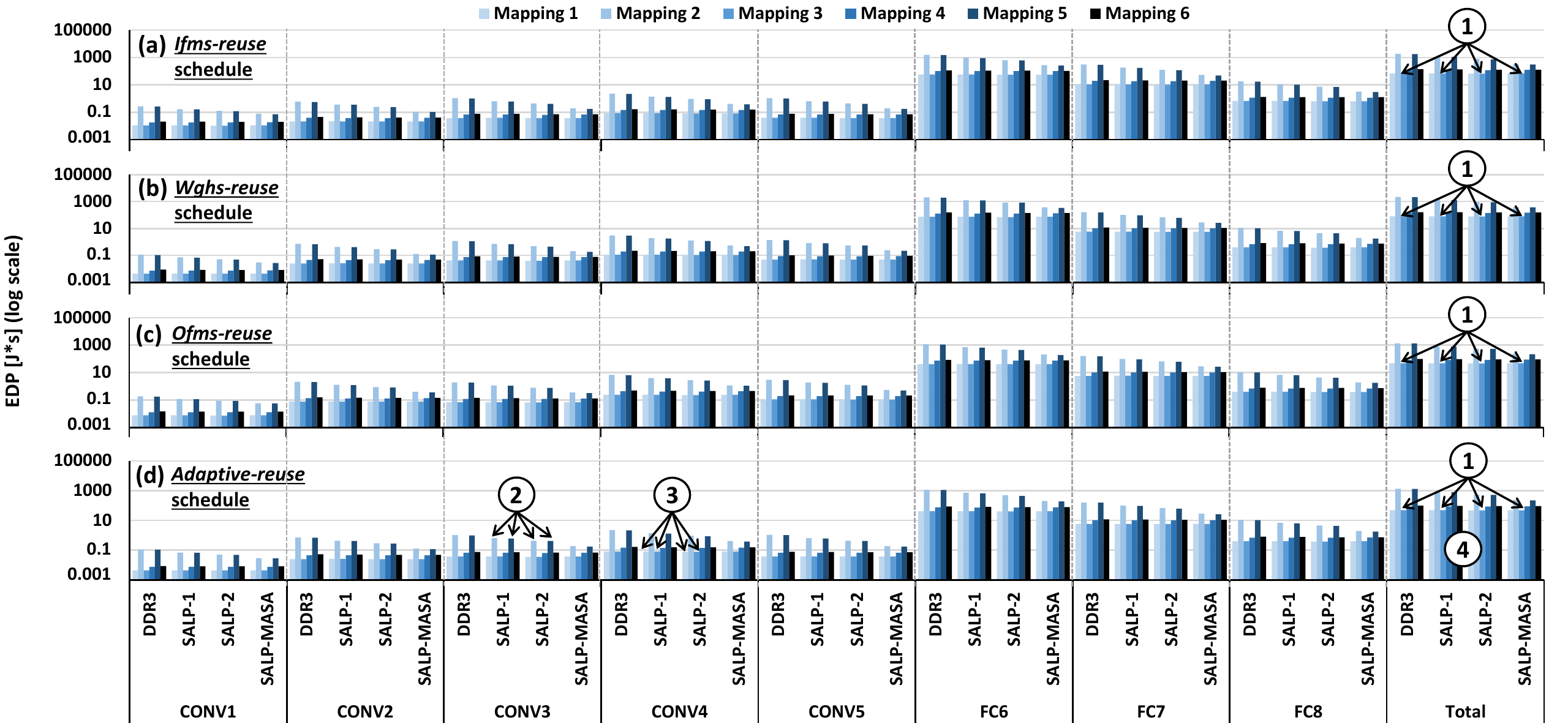}
	\vspace{-0.7cm}
	\caption{The EDP in AlexNet for different DRAM mapping policies across different DRAM architectures (DDR3, SALP-1, SALP-2, and SALP-MASA), while considering different scheduling schemes: (a) \textit{ifms}-reuse scheduling, (b) \textit{wghs}-reuse scheduling, (c) \textit{ofms}-reuse scheduling, and (d) adaptive reuse scheduling.
	}
	\label{Fig:Results_AlexNet_DifMap_ReuseAll}
	\vspace{-0.2cm}
\end{figure*}

\subsection{Comparisons of Different DRAM Mapping Policies}
\label{Sec:Results_DRAMmappings}

We evaluated the impact of different DRAM mapping policies and the results are presented in Fig. \ref{Fig:Results_AlexNet_DifMap_ReuseAll}.

\textbf{Key Observation-\circled{1}:} Our DRMap (Mapping-3) achieves the lowest EDP across different layers of the network, across different DRAM architectures, and across different scheduling schemes.
It indicates that the DRMap is the most effective DRAM mapping policy for different possible conditions.
According to Table \ref{Table:DRAMmaps}, DRMap (Mapping-3) orderly prioritizes to map the data to different columns in the same row (leading to row buffer hits in SALP and DDR3), to different banks in the same chip (exploiting bank-level parallelism in SALP and DDR3), to different subarrays in the same bank (exploiting subarray-level parallelism in SALP, but leading to row buffer conflicts in DDR3), and to different rows in the same subarray (leading to row buffer conflicts in SALP and DDR3).
\textit{Therefore, the DRMap is proven as a generic DRAM mapping policy that offers the lowest EDP.}
Moreover, different DRAM access scheduling schemes can make use of the DRMap, so that the CNN accelerators with different scheduling schemes can optimize their DRAM access latency and energy.
DRMap improves the EDP up to 96\% in DDR3, 94\% in SALP-1, 91\% in SALP-2, and 80\% in SALP-MASA, as compared to other mapping policies.

\textbf{Key Observation-\circled{2}:} Mapping-2 and Mapping-5 obtain worse EDPs (across different layers of the network, across different DRAM architectures, and across different scheduling schemes) than rest of the mapping policies.
The reason is that, Mapping-2 and Mapping-5 prioritize to map data across different subarrays in the same bank (exploiting subarray-level parallelism in SALP, but leading to row buffer conflicts in DDR3), that incurs higher latency and energy, as compared to row buffer hits and exploiting bank-level parallelism. 

\textbf{Key Observation-\circled{3}:} Mapping-1 and Mapping-3 obtain comparable EDPs.
The reason is that, Mapping-1 and Mapping-3 prioritize to map data across different columns in the same row (leading to row buffer hits in SALP and DDR3).
The difference comes when Mapping-1 proritizes to exploit subarray-level parallelism over bank-level parallelism, while Mapping-3 is the opposite.
From Fig. \ref{Fig:DRAM_HitMissConflict}, it is apparent that exploiting subarray-level parallelism incurs higher latency and energy than exploiting bank-level parallelism.

\subsection{Comparisons of Employing Different DRAM architectures}
\label{Sec:Results_DifferentDRAM}

%
In general, employing SALP architectures provides EDP improvements as compared to employing DDR3.
It is mainly due to latency and energy saving that are offered when exploiting subarray-level parallelism.
\textbf{Key Observation-\circled{4}:} For instance, if we consider \textit{adaptive-reuse} scheduling, EDP improvements achieved by employing SALP architectures as compared to DDR3 are:
\begin{itemize}[leftmargin=*]
    \item \textbf{For Mapping-1:} 0.59\% (SALP-1), 3.89\% (SALP-2), and 1.05\% (SALP-MASA).
    \item \textbf{For Mapping-2:} 29.18\% (SALP-1), 19.91\% (SALP-2), and 81.04\% (SALP-MASA).
    \item \textbf{For Mapping-3 (DRMap):} 0.6\% (SALP-1), 3.87\% (SALP-2), and 1.01\% (SALP-MASA).
    \item \textbf{For Mapping-4:} 0.71\% (SALP-1), 0.54\% (SALP-2), and 1.41\% (SALP-MASA).
    \item \textbf{For Mapping-5:} 29.67\% (SALP-1), 19.79\% (SALP-2), and 81.76\% (SALP-MASA).
    \item \textbf{For Mapping-6:} 3.15\% (SALP-1), 3.39\% (SALP-2), and 7.62\% (SALP-MASA).
\end{itemize}
The results show that employing SALP architectures is beneficial for improving energy-efficiency of DRAM accesses,
as along as an effective mapping policy like DRMap is employed.
The EDP of employing different DRAM architecture 
would be different, due to the different DRAM access energy and latency. 
However, since the internal organization of all DRAM architectures is similar (i.e., it is composed of channel, rank, chip, bank, subarray, row, and column as seen from top to bottom perspective), our DRMap can also be employed for all DRAM architectures to achieve the energy-efficient processing of convolutional neural networks in CNN accelerators.

\section{Conclusion}
\label{Sec:Conclusion}
In this paper, we present DRMap, a generic DRAM mapping policy that offers the lowest EDP of DRAM accesses for CNN accelerators, as compared to other mapping policies.
It is proven through an extensive design space exploration that study the latency and energy of different mapping policies, in different DRAM architectures as well as different layer partitioning and scheduling schemes. 
We expect that this work could enable further studies on energy-efficient CNN accelerators and help the existing CNN accelerators to optimize their DRAM access latency and energy. 

\section{Acknowledgment}
Authors acknowledge the scholarship granted by Indonesia Endowment Fund for Education (IEFE/LPDP), Ministry of Finance, Republic of Indonesia.

\bibliographystyle{IEEEtran}
\bibliography{bibliography}
\end{document}